\documentclass[12pt]{article}

\usepackage{multirow}
\usepackage[table,xcdraw,booktabs]{xcolor}
\usepackage{graphicx}
\usepackage{caption}
\usepackage{times}
\usepackage{booktabs}
\usepackage{amsfonts}
\usepackage{titlesec}
\usepackage{ulem}
\usepackage{natbib}
\usepackage{diagbox}
\usepackage[hidelinks]{hyperref}
\usepackage{geometry}
\makeatletter

\makeatletter

\begin{document} 

\title{Seismic Foundation Model (SFM): a new generation deep learning model in geophysics}

\author
{Hanlin Sheng$^{1}$, Xinming Wu$^{1\ast}$, Xu Si$^{1}$, Jintao Li$^{1}$, Sibo Zhang, and Xudong Duan\\
\normalsize{$^{1}$School of Earth and Space Sciences, University of Science and Technology of China,}\\
\normalsize{Hefei, China.}\\
\\
\normalsize{$^\ast$To whom correspondence should be addressed:}\\
\normalsize{E-mail: xinmwu@ustc.edu.cn}
}
\date{}




\maketitle

\begin{abstract}
While computer science has seen remarkable advancements in foundation models, which remain underexplored in geoscience. Addressing this gap, we introduce a workflow to develop geophysical foundation models, including data preparation, model pre-training, and adaption to downstream tasks. From 192 globally collected 3-D seismic volumes, we create a carefully curated dataset of 2,286,422 2-D seismic images. Fully using these unlabeled images, we employ the self-supervised learning to pre-train a Transformer-based Seismic Foundation Model (SFM) for producing all-purpose seismic features that work across various tasks and surveys. Through experiments on seismic facies classification, geobody identification, interpolation, denoising, and inversion, our pre-trained model demonstrates versatility, generalization, scalability, and superior performance over baseline models. Conclusively, we provide a foundation model and vast dataset to advance AI in geophysics, addressing challenges (poor generalization, lacking labels, and repetitive training for task-specified models) of applying AI in geophysics and paving the way for future innovations in geoscience.
\end{abstract}


\section*{Introduction}

Deep learning has demonstrated remarkable success in addressing complex challenges in Earth Sciences, encompassing atmospheric \citep{ham2019deep,kadow2020artificial,ravuri2021skilful,bi2023accurate}, oceanic \citep{andersson2021seasonal,bianco2019machine}, and solid earth domains \citep{lecun2015deep,bergen2019machine,reichstein2019deep}. In the field of solid earth, deep learning has facilitated the analysis of observational data to extract subsurface rock physics parameters and analyze physical processes, leading to a deeper understanding of Earth's internal structure and dynamics. Specifically, deep learning has contributed to solving various problems in the geophysical fields of seismology (phase picking \citep{ross2018p,zhu19phasenet,pardo2019seismic,liu2020deep,mousavi2020earthquake}, earthquake monitoring \citep{perol2018convolutional,ross2019phaselink,mousavi2019bayesian,rouet2020probing,zhu2022earthquake,zhu2022end,yang2022toward,wang2022predicting}, focal mechanism \citep{kuang2021real} and earthquake forecasting \citep{johnson2021laboratory,beroza2021machine}) and explorational geophysics (seismic data processing \citep{ovcharenko2019deep,park2020automatic,harsuko2022storseismic,mousavi2023applications}, interpolation \citep{wang2019deepI,chai2020deep,kaur2021seismic}, interpretation \citep{wu2019faultseg3d,shi2019saltseg,pham2019automatic,geng2020automated,li2021neural}, and inversion \citep{wu2019inversionnet,yang2019deep,chen2020seismic,li2023self,shi2023semi}). These efforts have shown promising results, but mostly following the way of training a specific model for each task. Such a specific model is often only applicable to a particular task, a small datasets with specific patterns or acquired at specific surveys, which makes us in dire need of a model with better generalization.

The recent rise of fundamental models is promising to address the poor generalization problem\citep{li2023machine}. The concept of a Foundation Model, a model pre-trained on large-scale data in a self-supervised or semi-supervised manner that can be adapted for various downstream tasks \citep{bommasani2021opportunities}, has triggered a new revolution in the field of artificial intelligence. These models are trained once on extensive datasets and subsequently applied to address a wide range of related tasks. The recent emergence of models with large number of parameters, like GPT-4 \citep{openai2023gpt4}, Pathways Language Model (PaLM, \citep{driess2023palm}), Vision Transformer-22B \citep{dehghani2023scaling}, MAE \citep{He_2022_CVPR}, VideoMAE V2 \citep{wang2023videomae}, CLIP \citep{radford2021learning}, Segment Anything Model (SAM, \citep{kirillov2023segment}), and DINO V2 \citep{oquab2023dinov2}, has highlighted the potent feature extraction capabilities. Such models can tackle previously challenging problems and even excel at unforeseen tasks. In the field of geophysics, a demand for this transformative potential exists, but direct application of these models is hindered by differnet data statistical properties, disparities in spatiotemporal resolutions, and significant variations in physical meanings \citep{reichstein2019deep}. Consequently, a geoscience-specific foundation model is required \citep{li2023machine}.

\begin{figure}[!bt]%
\centering
\includegraphics[width=\textwidth]{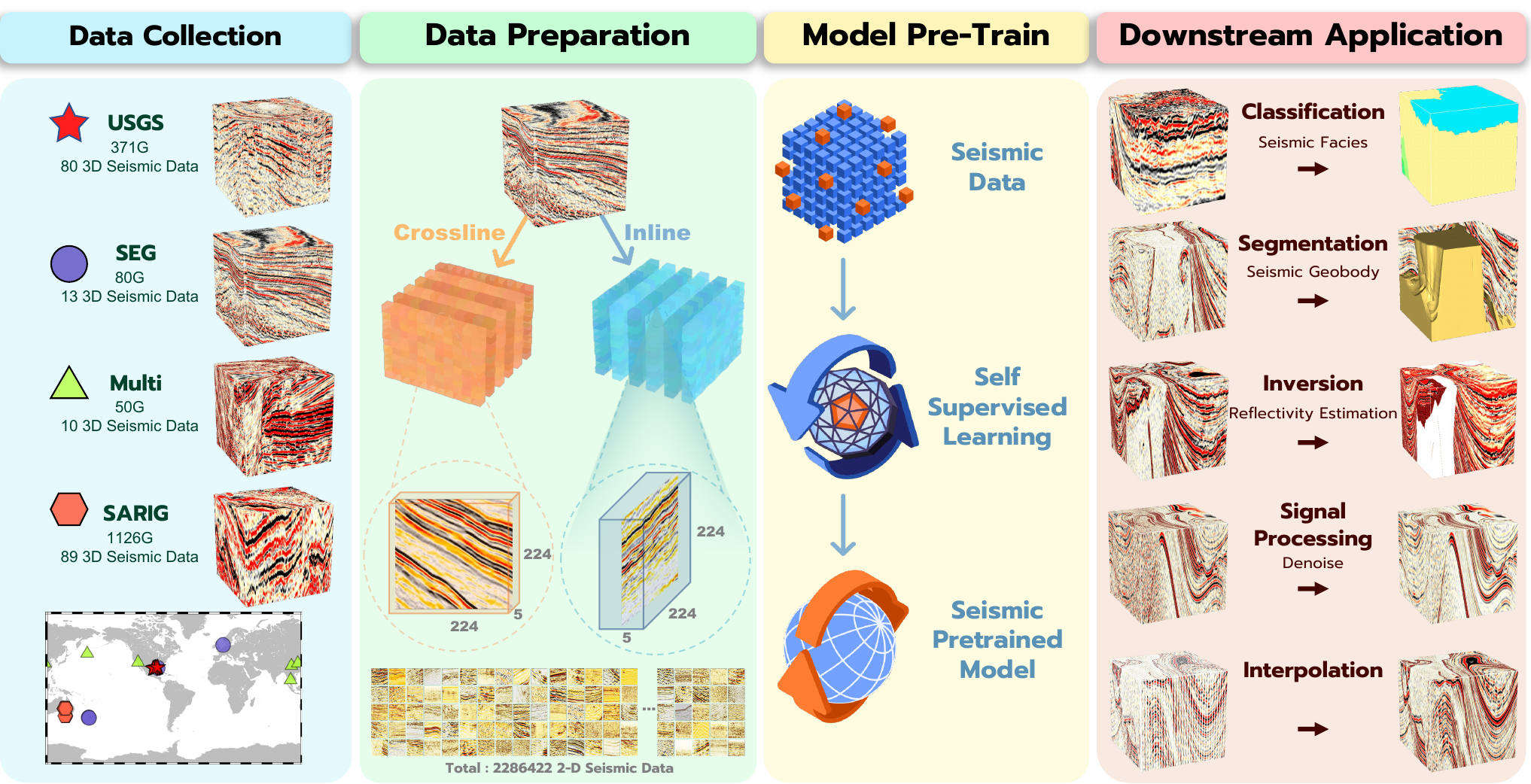}
\caption{
Four main stages of the workflow for developing our seismic foundation model: diverse raw data collection, training data preparation, model pre-training, and downstream task application. First, a diverse seismic dataset is collected worldwide from 192 3-D surveys, from which 2286422 2-D seismic data are carefully chosen for pre-training. The foundational model is further constructed and trained through a self-supervised pre-training strategy on the vast amount of unlabelled seismic data. The versatility of the pre-trained model is finally demonstrated in downstream applications such as classification (i.e. seismic facies), segmentation (i.e. seismic geobody), inversion (i.e. reflectivity estimation), signal processing (i.e. denoise), and interpolation.}
\label{fig1}
\end{figure}

To develop such a model, extensive and diverse geophysics data must be collected, followed by training methodologies tailored to the characteristics of geophysical data. Subsequently, the model's effectiveness should be validated across various geophysical tasks. In this context, we present a whole workflow to build a geophysical foundational model by using seismic data as a case study (Fig. \ref{fig1}). We collected a global set of 192 3-D seismic datasets and meticulously prepared them for effective training. Given the lack of labels in seismic data and the need of global context understanding for most seismic data processing and interpretation tasks, we selected a generative self-supervised training strategy combined with the Transformer architecture for developing our seismic foundation model. We applied the resulted foundation model to downstream tasks such as classification (i.e. seismic facies), segmentation (i.e. seismic geobody), inversion (i.e. reflectivity estimation), signal processing (i.e. denoise), and interpolation. Across all the tasks, our foundation model demonstrated superior performance compared to task-specifically trained networks, highlighting the effectiveness of our provided dataset and training workflow.

\section*{Results}

\subsection*{Data Collection}

To train our foundational model effectively, a substantial and diverse dataset with rich features is essential. We collected diverse 3-D migrated seismic data from the sources of United States Geological Survey (USGS, \citep{USGS}), South Australian Resources Information Gateway (SARIG, \citep{SARIG}), Society of Exploration Geophysicists (SEG, \citep{SEG}), and so on. The 3-D seismic data obtained from USGS primarily located in the Gulf of Mexico Basin, United States. We gathered a total of 80 datasets from this source, amounting to around 371 gigabytes (G). SARIG's data encompassed the southern region of Australia, where we collected 89 3-D datasets, totaling 1126 G. The SEG data consisted of publicly available seismic surveys from various locations worldwide, from which we collected a total of 13 3-D seismic volumes, with a size of approximately 80 G. Additionally, we incorporated data from various other sources, resulting in a collection of 10 datasets totaling 50 G. In total, we amassed 192 3-D migrated seismic datasets acquired in the global areas of Middle America, Southern Australia, Southeast Asia, and Northern Europe, amounting to 1562 G of data. These massive datasets sourced from worldwide encompassed a wide range of subsurface geologic features. These features included various types of faults (e.g., normal, reverse, strike-slip), varying degrees of folding, geobodies of different sizes and locations, as well as unconformities exhibiting different combinations and types. These representative geological characteristics provide valuable insights for analyzing and interpreting the seismic data.

\subsection*{Training data preparation}

After acquiring the seismic datasets, data preprocessing is necessary to meet the requirements of training a foundation model. Considering the prohibitive GPU memory and computational demands of training a large-scale foundation model using 3-D data, we selectively extracted a diverse 2-D training dataset from the collected 3-D dataset to train our foundation model. As shown in the second column in Fig. \ref{fig1}, we performed 2-D seismic slicing on each seismic dataset in both the inline and crossline directions. The 3-D data were segmented into smaller blocks of size 224 (vertical dimension) $\times$ 224 (inline/crossline direction) $\times$ 5 (crossline/inline direction). From the five available images in each block, we selected the one with the greatest difference to the previously selected sample to be added to the training dataset. In this way, we extracted one 2-D sample for every five inline or crossline slices, which was helpful to reduce the redundancy of features between the extracted samples. Additionally, we introduced overlapping regions between two adjacent blocks in the vertical direction, effectively increasing the number of training samples. A subset of these samples from our dataset was displayed at the bottom of the second column of Fig. \ref{fig1}, illustrating the diverse structural features contained in the samples.

\subsection*{Model Training}\label{subsec3}

\begin{figure}[htb!]%
\centering
\includegraphics[width=\textwidth]{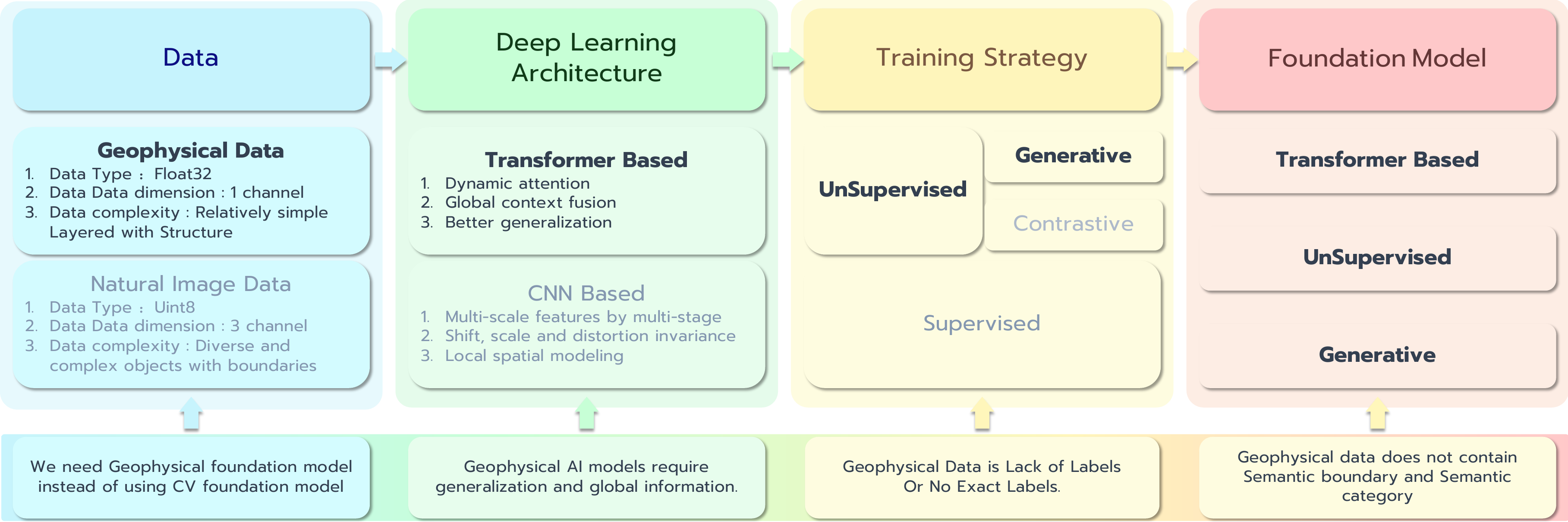}
\caption{Our underlying logic of choosing the network architecture and pre-training strategy for the seismic foundation model. We illustrate the contrast between seismic data's distinctive characteristics and those of natural images, driving the selection of generative unsupervised learning and the Transformer architecture as a domain-specific strategy for developing our foundation model.
}\label{fig:supp_figure2}
\end{figure}

Selecting a suitable approach for training a seismic foundational model necessitates a careful consideration of the uniqueness of seismic images (see Fig. \ref{fig:supp_figure2}). Seismic images are typically represented in Float32 precision, which is higher than the Uint8 precision often used for natural images. Additionally, seismic data comprise a single channel, whereas natural images typically exhibit RGB three-channel structures. From a visual standpoint, seismic images contain relatively simple layered structures, while natural images consist of more diverse and complex objects with well-defined boundaries. In addition to the above differences beyond the composition of the data, the physical significance and the statistics of the data are also considerably distinct from each other. These discrepancies imply a gap when directly applying computer vision-based foundational models to the field of geophysics, necessitating the development of domain-specific foundational models. Given that most seismic data lack annotations or ground truth, we considered using the unsupervised learning as the preferred training strategy to make it possible to pre-train our foundation model on a large amount of unlabeled datasets. Unsupervised learning consists of two major categories of generative and contrastive learning. Generative unsupervised learning involves predicting a portion of raw data from the remaining portion, while contrastive unsupervised learning entails creating positive and negative perspectives of the input, aligning the positives, and distinguishing the negatives \citep{wang2023scientific}. Unlike natural images where object categories (e.g., cats versus dogs) can be readily defined as negative samples, different categories within seismic data may lack distinct boundaries or differentiation. Seismic data from diverse samples could share geological significance, such as belonging to the same stratigraphic unit or sedimentary/tectonic environment. Besides, the categories within seismic data can change according to the purpose of collecting and analyzing the data \citep{reichstein2019deep}. Thus, we opted for a generative unsupervised learning approach for our training strategy. This choice was aligned with the inherent complexities of seismic data, allowing us to uncover meaningful patterns and structures within the subsurface.

Regarding the architecture, the majority of foundation models currently adopt the Transformer framework \citep{vaswani2017attention}, owing to its dynamic attention mechanism tailored to different input data and its ability to capture global context information. More importantly, Transformer exhibits great generalization properties, which are crucial for the application of deep learning in seismic data. As a contrasting option, CNN-based network architectures \citep{ronneberger2015u,chen2017rethinking} can obtain multi-scale features through multiple stages and demonstrate shift, scale, and distortion invariance due to the convolutional algorithms. CNN also excels in local spatial modeling. However, given the seismic data's demand for global information and improved generalization, we chose the Transformer architecture to build our foundational model. In summary, considering the characteristics of seismic data, we opted for the combination of generative unsupervised learning and the Transformer architecture as our strategy for developing our foundation model. Upon adopting this strategy, various methods are available for its implementation, and in this study, we chose the Masked Autoencoders (MAE \citep{He_2022_CVPR}) method (Supplementary Fig. \ref{fig:supp_figure1}). This choice was motivated by the fact that the masking strategy in MAE significantly reduces the required memory and training time. After completing the pre-training of the foundation model, we extracted the Encoder Module as the seismic foundation model (SFM). The SFM was then utilized to extract features and fed them into a simplified decoder network for further application. More details about the training process were included in the supplementary file.

\subsection*{Application}\label{subsec4}

Following the completion of model training, we have collected diverse datasets to evaluate the performance of our foundation model. For segmentation tasks, we assessed the model's capabilities in classification (i.e. seismic facies) and segmentation (i.e. seismic geobody). Additionally, we tested the model's performance in regression tasks, specifically in signal processing (i.e. denoising), inversion (i.e. reflectivity estimation), and interpolation.

\begin{figure}[!hbtp]
\centering
\includegraphics[width=0.85\textwidth]{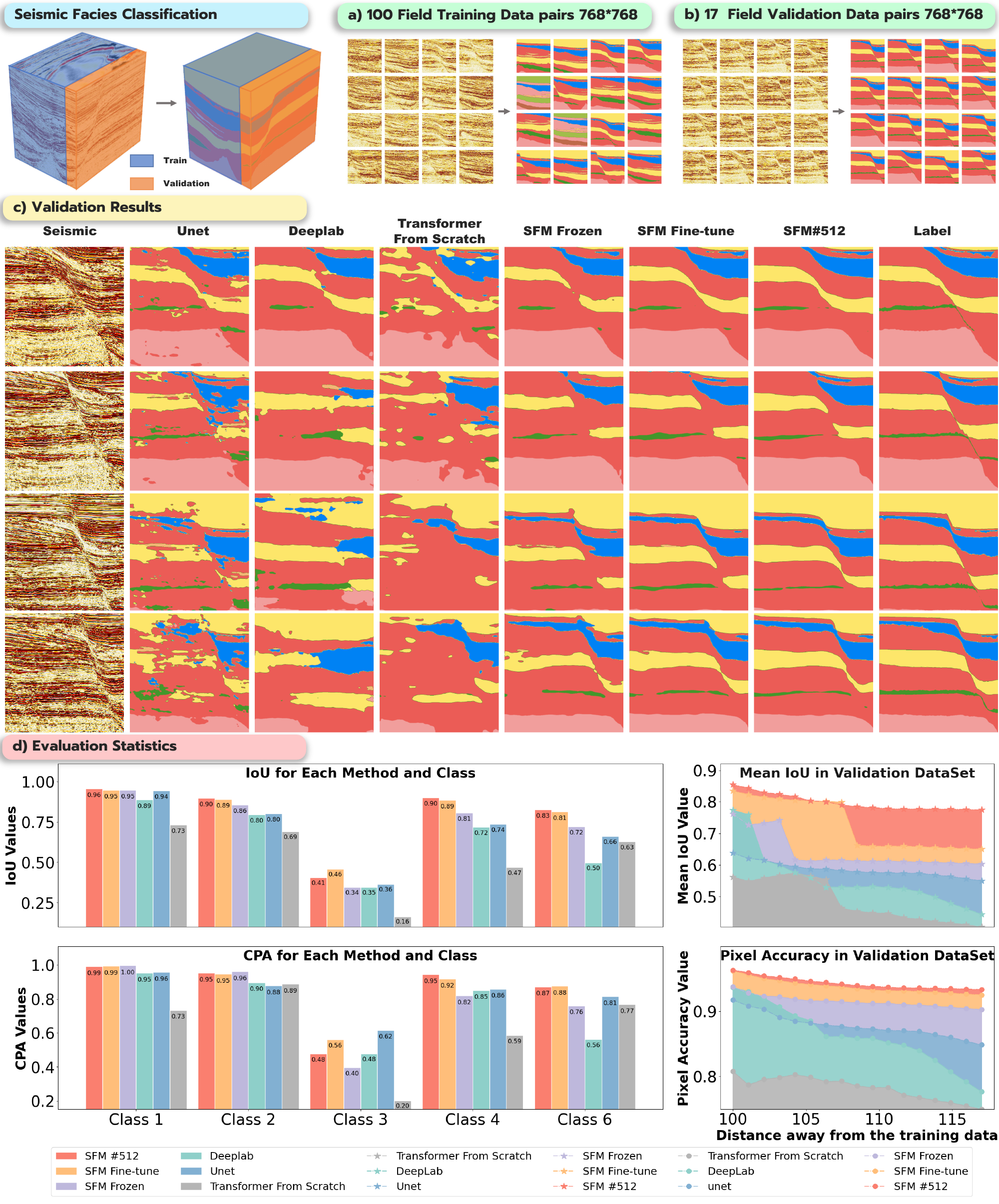}
\caption{
Seismic Facies Classification Task: Leveraging the Facies Identification Challenge seismic facies dataset, we employed the first 500 crossline slices for training and the subsequent 90 crossline slices for validation. Considering the strong similarity between adjacent slices, we selected one every five slices, ultimately forming a training dataset with 100 slices (a) and a validation dataset with 17 slices (b). In the validation dataset, we present results obtained from Unet, Deeplab, Transformer From Scratch, SFM Frozen, SFM Fine-tune, and SFM \# 512 models. The predictions using the first three models are not as effective as desired when compared to the ground truth. However, based on our pre-trained foundation model, SFM \# 512, SFM Frozen and SFM Fine-tune outperform the Unet and Deeplab methods, showcasing the significant impact of our pre-trained model in enhancing model performance (c). The bar charts displayed the IoU and CPA for each class. Notably, the SFM Fine-tune and SFM Frozen methods consistently demonstrate stable and superior performance across all classes (d). We also show the mean IoU and Accuracy of the models on the validation seismic slices that are arranged in terms of distance from the training dataset (lower-right panels). We observe that the SFM Fine-tune and SFM Frozen models' performance remains more stable as they move further away from the training data, indicating better generalization.
}\label{fig2}
\end{figure}

\paragraph{Seismic Facies Classification}
A seismic facies unit refers to a delineated seismic unit comprising clusters of reflections exhibiting distinct characteristics from adjacent facies units. Accurately identifying seismic facies is essential for understanding ancient stratigraphic and structural changes. However, manual delineation of seismic facies can be challenging, requiring extensive geological background knowledge and physical parameters inferred from seismic data. We collected data from the competition ``Facies Identification Challenge: 3-D image interpretation by machine learning techniques", jointly organized by AIcrowd and SEAM. The dataset provided for this competition consists of a 3-D seismic image from the publicly available ``Parihaka" seismic survey \citep{SEG}. Experts have meticulously interpreted each point in the image and classified it into one of six different facies (see Fig. \ref{fig2}).

We partitioned 590 crosslines of the seismic volume into a 500:90 ratio, with 500 serving as the training dataset (shown in blue) and 90 as the validation dataset (shown in orange). Considering that adjacent seismic images exhibit similar features, we randomly selected representative seismic images from every five adjacent seismic images in both the training and validation datasets. Consequently, we obtained 100 training samples (Fig. \ref{fig2}a) and 17 validation samples(Fig. \ref{fig2}b).

We compared our approach, referred to as ``Seismic Foundation Model (SFM)" with two commonly used segmentation models: Unet \citep{ronneberger2015u} and Deeplab \citep{chen2017rethinking}. Additionally, we conducted experiments with variations of our Transformer model, including ``Transformer From Scratch" (the same architecture without pre-trained parameters), ``SFM Frozen" (our model architecture with pre-trained parameters, with only Decoder parameters training), and ``SFM Fine-tune" (our model architecture with pre-trained parameters, with full fine-tuning). 

In the seismic facies classification task, geophysicists are particularly interested in obtaining more continuous seismic facies results and emphasizing the delineation of contact relationships between different facies. The experimental results for seismic facies, as shown in Fig. \ref{fig2}c, indicated that the predictions using the Unet, Deeplab, and Transformer From Scratch were not as effective as desired when compared to the labeled ground truth. However, with the pre-trained seismic foundational model, both SFM Frozen and SFM Fine-tune outperformed the Unet and Deeplab methods, showcasing the significant impact of our pre-trained model in achieving better performance. To provide a comprehensive comparison of the results, we introduced mean Intersection over Union (mIoU) and Class Pixel Accuracy (CPA) as metrics to evaluate the performance of different methods. From the overall performance on the validation dataset (as shown in Table \ref{tab1}), the SFM Fine-tune and SFM Frozen models outperformed other ones. In the Evaluation section of Fig. \ref{fig2}d, the bar charts displayed the IoU and CPA for each class. Notably, the SFM Fine-tune and SFM Frozen models consistently demonstrate stable and superior performance across all classes.

As shown in the right panels in Fig. \ref{fig2}d, we also evaluated the generalization capabilities of different models on the validation slices that were increasingly distant from the training samples. As the validation seismic slices gradually move away from the training samples, the differences between these slices and the training samples also increase. Stable seismic facies prediction on the slices requires the model to have strong generalization capability. The area under the curve in the line plot (right panels in Fig. \ref{fig2}d) demonstrated that the SFM Fine-tune and SFM Frozen models, in relation to other ones, exhibited a slower performance decline with distance. This indicates that our approach achieves more stable extrapolation of interpretative results, which is crucial in assisting experts during seismic facies interpretation and highlights the improved generalization.

\paragraph{Seismic Geobody Identification}

\begin{figure}[!htbp]%
\centering
\includegraphics[width=0.9\textwidth]{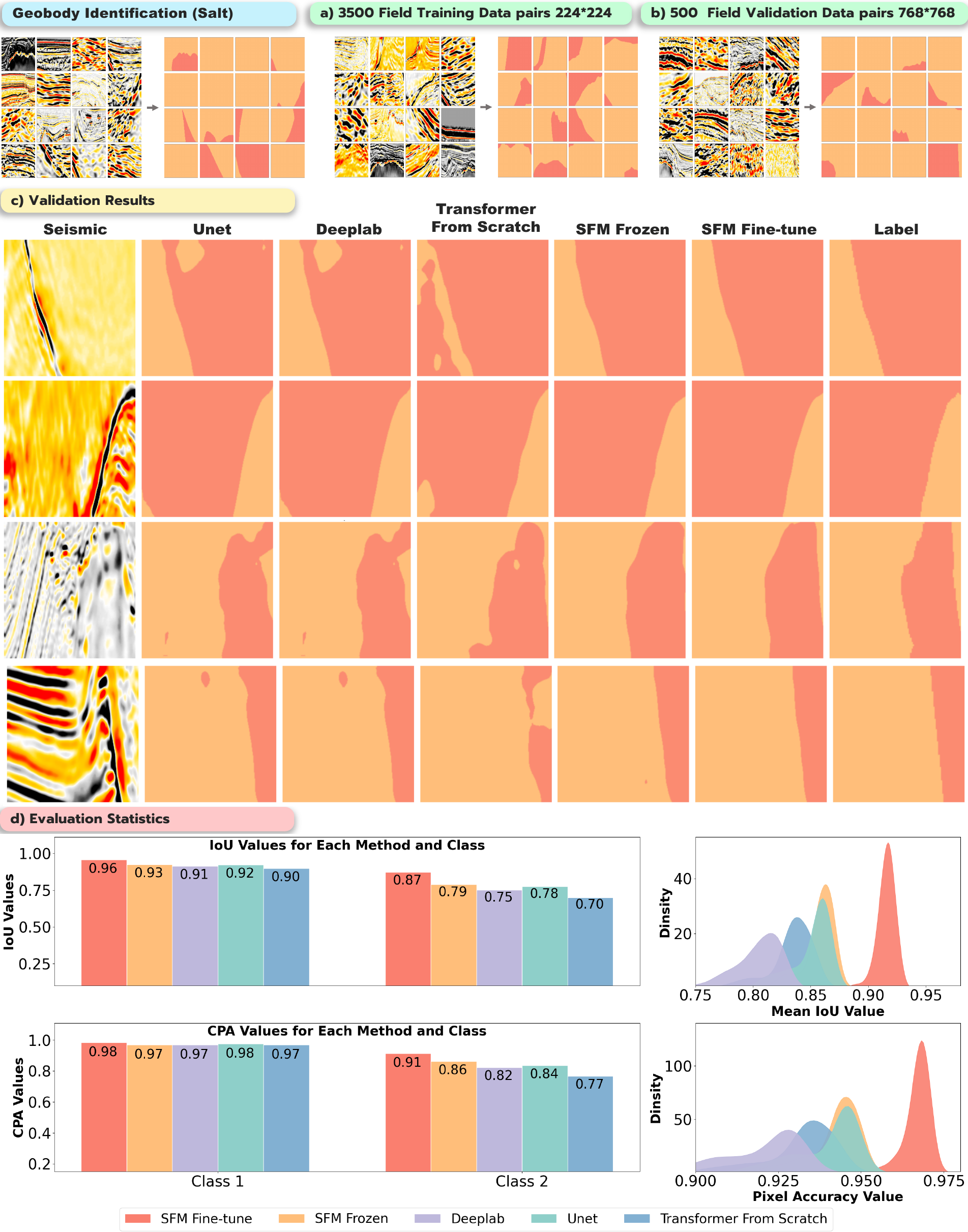}
\caption{Seismic Geobody (Salt) Identification Task: Utilizing the TGS Salt Identification Challenge seismic salt dataset, we allocated 3500 data for training (a) and reserved the remaining 500 data for validation (b).In this dataset, we test the Unet, Deeplab, Transformer From Scratch, SFM Frozen, and SFM Fine-tune models (c). The predictions from the Unet, Deeplab, and Transformer From Scratch models exhibit outliers and holes that are mis-predicted. Conversely, the SFM Frozen and SFM Fine-tune models displayed a more reasonable pattern with continuous predictions, aligning better with the labels. The bar charts illustrate the IoU and CPA for each class (d). Notably, the SFM Fine-tune and SFM Frozen models consistently exhibit robust and superior performance across all classes. Additionally, we present statistical summaries of mean IoU and Accuracy for various methods on the validation dataset, where the SFM Fine-tune and SFM Frozen models tend to concentrate on higher IoU and Accuracy, suggesting a more favorable performance of our proposed methods.
}\label{fig3}
\end{figure}

Another crucial task in seismic interpretation is the recognition of geological bodies, known as geobodies. The unique subsurface features, such as salt, karst, and channels, often indicate different rock types underground, provide insights into specific tectonic phenomena, and reveal the evolution of geological structures. We took the task of salt recognition as a representative example to demonstrate the performance of our SFM on the tasks of seismic geobody identification. We collected data from the Kaggle competition ``TGS Salt Identification Challenge". The dataset consisted of 4000 samples, each of size 101$\times$101. To accommodate our model's input size, we interpolated the data to seismic data of size 224$\times$224. Among these samples, 3500 were assigned to the training dataset (Fig. \ref{fig3}a), while the remaining 500 samples served as the validation dataset (Fig. \ref{fig3}b). 

Similarly, we compared the Unet, Deeplab, Transformer From Scratch, SFM Frozen, and SFM Fine-tune models. From the overall performance on the validation dataset (as shown in Table \ref{tab1}) and the visualizations of the predictions on the validation dataset (Fig. \ref{fig3}c), it was evident that both the SFM Fine-tune and SFM Frozen models consistently exhibited superior IoU and CPA for the two classes. We also showed the statistics of mean IoU and Pixel Accuracy across all validation samples (Fig. \ref{fig3}d). The SFM Fine-tune and SFM Frozen models outperformed the other ones in the majority of the validation dataset (with density concentrated in the high mean IoU and Pixel Accuracy region). 


\newcommand{\tabincell}[2]{\begin{tabular}{@{}#1@{}}#2\end{tabular}}

\begin{table}[!htbp]
\begin{center}
\begin{minipage}{\textwidth}
\caption{
Segment Task Performance
}\label{tab1}
\begin{tabular*}{\textwidth}{@{\extracolsep{\fill}}lcccc@{\extracolsep{\fill}}}
\toprule
& \multicolumn{2}{@{}c@{}}{Seismic Facies Classification} & \multicolumn{2}{@{}c@{}}{Seismic Geobody Identification} \\\cmidrule{2-3}\cmidrule{4-5}%
Method & mean IoU &  Pixel Accuracy & mean IoU &   Pixel Accuracy  \\
\midrule
Unet  									& 0.5851 & 0.8779  & 0.8497 & 0.9396\\
Deeplab  								& 0.5564  & 0.8618 & 0.8325 & 0.9317\\
 \tabincell{l}{Transformer \\From Scratch}  			& 0.4858 & 0.7822  & 0.7989 & 0.9178\\
\midrule
SFM Frozen            & 0.6417 & 0.9148 & 0.8577 &  0.9424\\
SFM Fine-tune         & 0.7294 & 0.9377  & \textbf{0.9145} & \textbf{0.9667}\\
SFM Fine-tune 512  & \textbf{0.7980}  &  \textbf{0.9430}   & -  &-  \\
\bottomrule

\end{tabular*}
\footnotetext{
Note: Bold values represent the best performance.}
\end{minipage}
\end{center}
\end{table}

\paragraph{Inversion (Reflectivity Estimation)}

In the field of geophysics, obtaining accurate subsurface physical parameters remains a challenging task. Extracting real-world rock properties from borehole logging is both costly and impractical for large-scale applications. Hence, geophysicists resort to various feasible methods for signal acquisition. Once signals are acquired, geophysicists undergo signal processing and inversion techniques to reflect the desired subsurface information and rock properties, such as lithology. Seismic inversion is one such method, mapping geophysical data to physical models that help understand subsurface physical properties like reflectivity, impedance, and water saturation. These inverted subsurface physical models can be used for disaster early warning, energy exploration, and the development of underground spaces in smart cities.

\begin{figure}[!hbtp]%
\centering
\includegraphics[width=\textwidth]{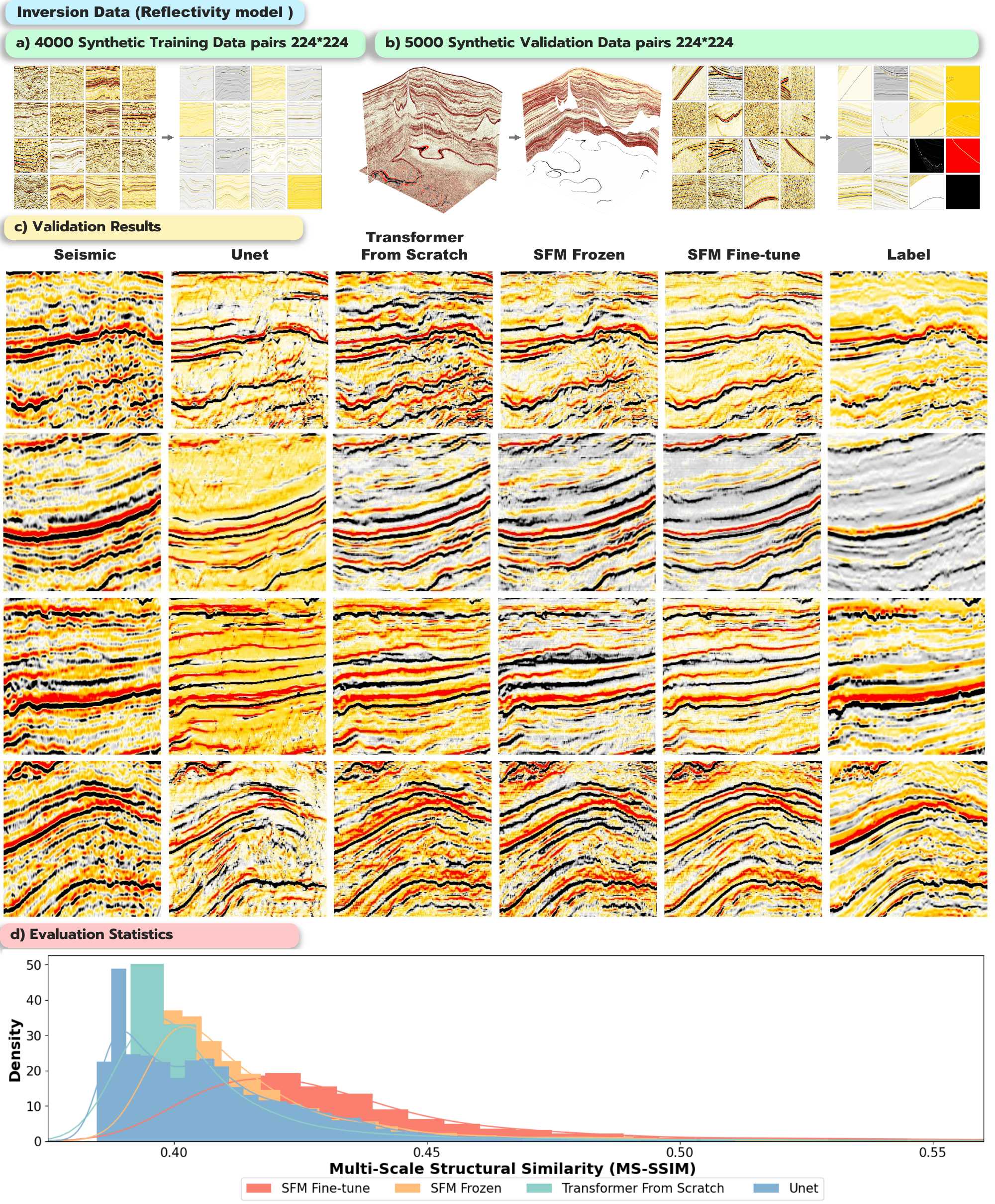}
\caption{Seismic Inversion (Reflectivity Model) Task: We synthesized 4000 data for training (a) and employed 5000 validation samples, extracted from a more realistic SEAM phase I model, for validation (b). For this task, we compared the performance of the Unet, Transformer From Scratch, SFM Frozen, and SFM Fine-tune models(c). The predictions of the Unet model exhibit erratic features in low signal-to-noise ratio areas and lack continuity in reflectors. Conversely, other three models displayed more continuous reflectors. The statistical summaries of MS-SSIM between predictions and labels (d) reveal that the SFM Frozen and SFM Fine-tune models predominantly converge in regions of higher structural similarity, signifying greater similarity in the predicted reflectivity models.
}\label{fig4}
\end{figure}

We tested our SFM's application in the task of reflectivity model inversion, which served as a representative example of inversion problems. Due to the lack of real labels, we simulated a reflectivity-seismic dataset as the training dataset and utilized the 3-D seismic volume and the corresponding reflectivity model from the SEAM Phase I model as the validation dataset. Specifically, as shown in Fig. \ref{fig4}, we performed geological forward modeling to generate 2200 seismic reflectivity models of size 224$\times$224 \citep{wu2020building} and simulated their corresponding seismic data as the training dataset (Fig. \ref{fig4}a). For the validation dataset, we segmented the SEAM Phase I model (Fig. \ref{fig4}b) by using the same method in \textbf{Data preparation} section. Finally, we obtained a total of 5000 pairs of 224$\times$224 seismic data for the validation dataset (Fig. \ref{fig4}b).

Again, we compared the Unet, Transformer From Scratch, SFM Frozen, and SFM Fine-tune models. As shown in the validation results (Fig. \ref{fig4}c), the Unet model exhibited shortcomings in some low signal-to-noise ratio (SNR) data, presenting abrupt short and cluttered predictions of the reflections. Such performance was deemed unacceptable in practice. On the other hand, the Transformer-based methods demonstrated superior performance in this aspect. To provide a comprehensive comparison, we introduced Multi-Scale Structural Similarity (MS-SSIM) as a metric to evaluate the model performance. Since the predictions of all the networks were relative reflectivity models, evaluating absolute reflectivity values was not appropriate. Therefore, we only used MS-SSIM as the evaluation parameter. The overall performance on the validation dataset of 5000 samples (as shown in Table \ref{tab2}) indicated that both SFM Fine-tune and SFM Frozen, which are associated with our seismic foundational model, outperformed other models. Furthermore, the MS-SSIM distribution plot in Fig. \ref{fig4}d showed that SFM Fine-tune and SFM Frozen models predominantly concentrated in regions with higher structural similarity. By training on synthetic data, our method can also be generalized to field data (Supplementary Figure \ref{fig:supp_figure3}). These results demonstrated that our foundation model leveraged synthetic data to incorporate geophysicists' understanding of physical processes for inversion results. 

\paragraph{Denoise}

\begin{figure}[!hbtp]%
\centering
\includegraphics[width=0.85\textwidth]{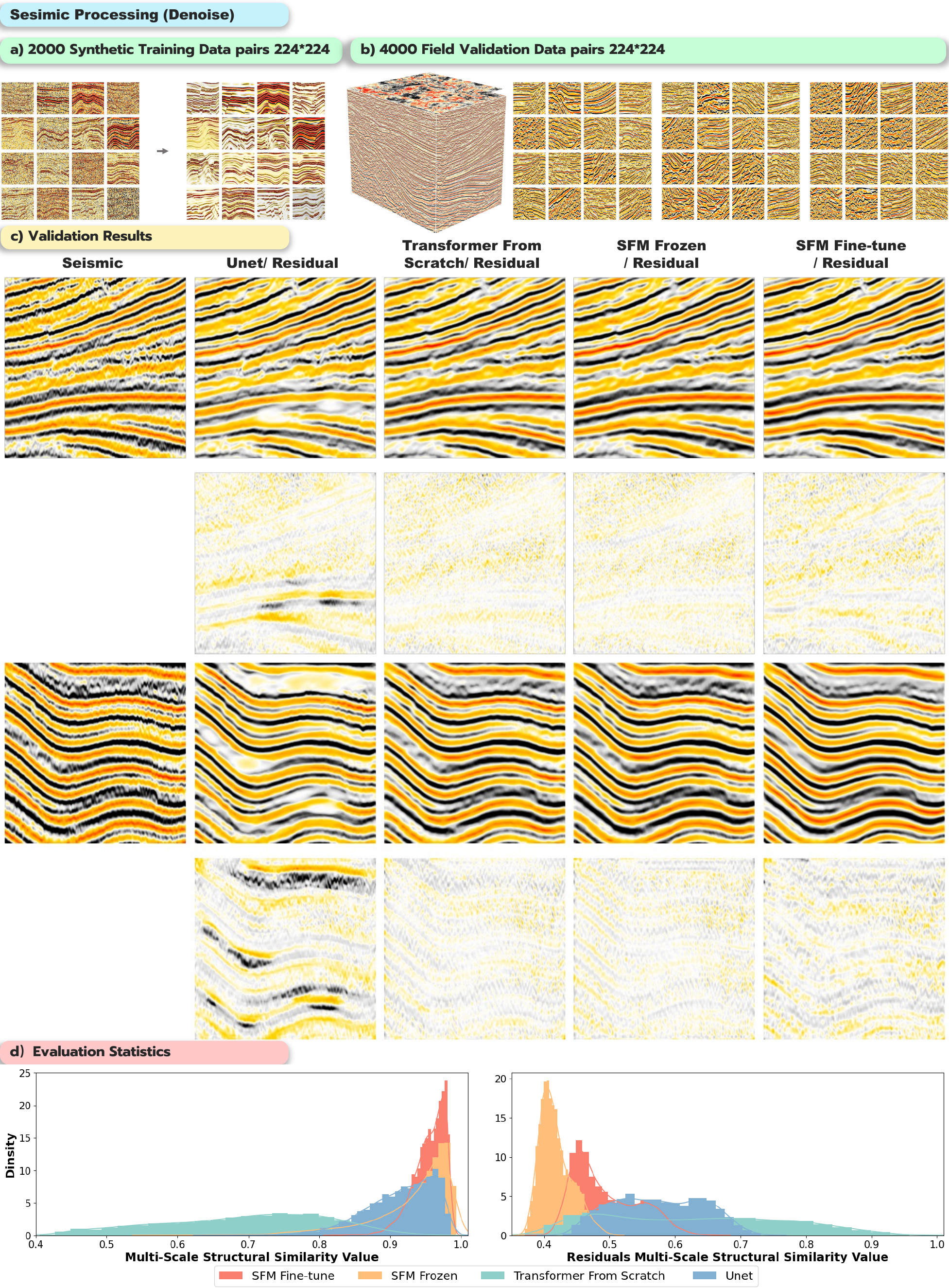}
\caption{
Seismic Signal Processing Data (Denoise) Task: We synthesized 2000 data for training (a)and extracted 4000 profiles from a field seismic data for validation (b). On this dataset, we conducted comparative testing of the Unet, Transformer From Scratch, SFM Frozen, and SFM Fine-tune models (c). Predictions from the Unet models exhibit an over-removal of valid signals, resulting in discontinuities and information loss in structures. In contrast, predictions from the other three models demonstrate an enhanced data signal-to-noise ratio, with removed noise containing minimal strong energy signals. The statistical graphs showcase MS-SSIM between predictions and original data, as well as MS-SSIM between residuals and original data (d). Notably, predictions from the SFM Frozen and SFM Fine-tune models align predominantly in regions of higher structural similarity, with residuals focusing on areas of lower structural similarity. This suggests that our predicted results closely resemble the original data in terms of structure, and the removed noise contains fewer valid signals.
}\label{fig5}
\end{figure}

Signal processing is another key research focus in the field of geophysics. Due to variations in acquisition devices and processing techniques, geophysicists often encounter subsurface images with different signal-to-noise ratios (SNRs). The SNR directly impacts our ability to extract crucial structural information and lithological variations. Various types of interference noise pose challenges, including random noise, radio industry interference, low-frequency noise, ground-roll noise, and multiple reflections \citep{mousavi2022deep}. The objective of noise removal is to eliminate such interference signals and highlight the essential information. However, in field data, distinguishing between noise and valuable signals might be difficult. Hence, for this task, we employed forward modeling to synthesize the dataset. We generated 2000 noise-free seismic data and subsequently added random noise to obtain seismic data with noise. Each pair of data before and after noise addition formed a training sample pair (Fig. \ref{fig5}a). For the validation dataset, we opted to evaluate the model's performance on field data. Thus, we selected a 3-D dataset (Fig. \ref{fig5}b) and segmented it into 2-D 224$\times$224 seismic data by using the method mentioned in \textbf{Data preparation} section, resulting in a total of 4000 224$\times$224 profiles for the validation dataset (Fig. \ref{fig5}b).

In the noise removal task dataset, we also compared the performance of Unet, Transformer From Scratch, SFM Frozen, and SFM Fine-tune models. The results (Fig. \ref{fig5}c) on the validation dataset demonstrated the difference between the predicted results and the original noisy data. Ideally, this removal noise should not contain any useful signal, and the predicted denoise result should retain the same structure as the original data. We observed that the Unet model tended to remove signals during the denoising process, leading to structural losses in some areas. This is an undesirable outcome in denoise methods. Transformer-based methods, on the other hand, performed relatively better in preserving the structure. Due to the lack of true labels without noise, to better describe the comparison results, we not only compared the MS-SSIM between the predicted results and the original seismic data but also between the residuals and the original seismic data. We expected the denoised images to resemble the original images more while making the residuals less similar to the original seismic images. The overall performance of the 4000 samples in the validation dataset (as shown in Table \ref{tab2}) indicated that both SFM Fine-tune and SFM Frozen methods performed well. Moreover, from the MS-SSIM distribution of the predicted results and the residuals in Fig. \ref{fig5}d, we can observe that SFM Fine-tune and SFM Frozen methods primarily focused on regions with higher structural similarity for the predicted results and regions with lower structural similarity for the residuals, which aligned with our expectations for denoise. 

\paragraph{Interpolation}

\begin{figure}[!hbtp]%
\centering
\includegraphics[width=0.87\textwidth]{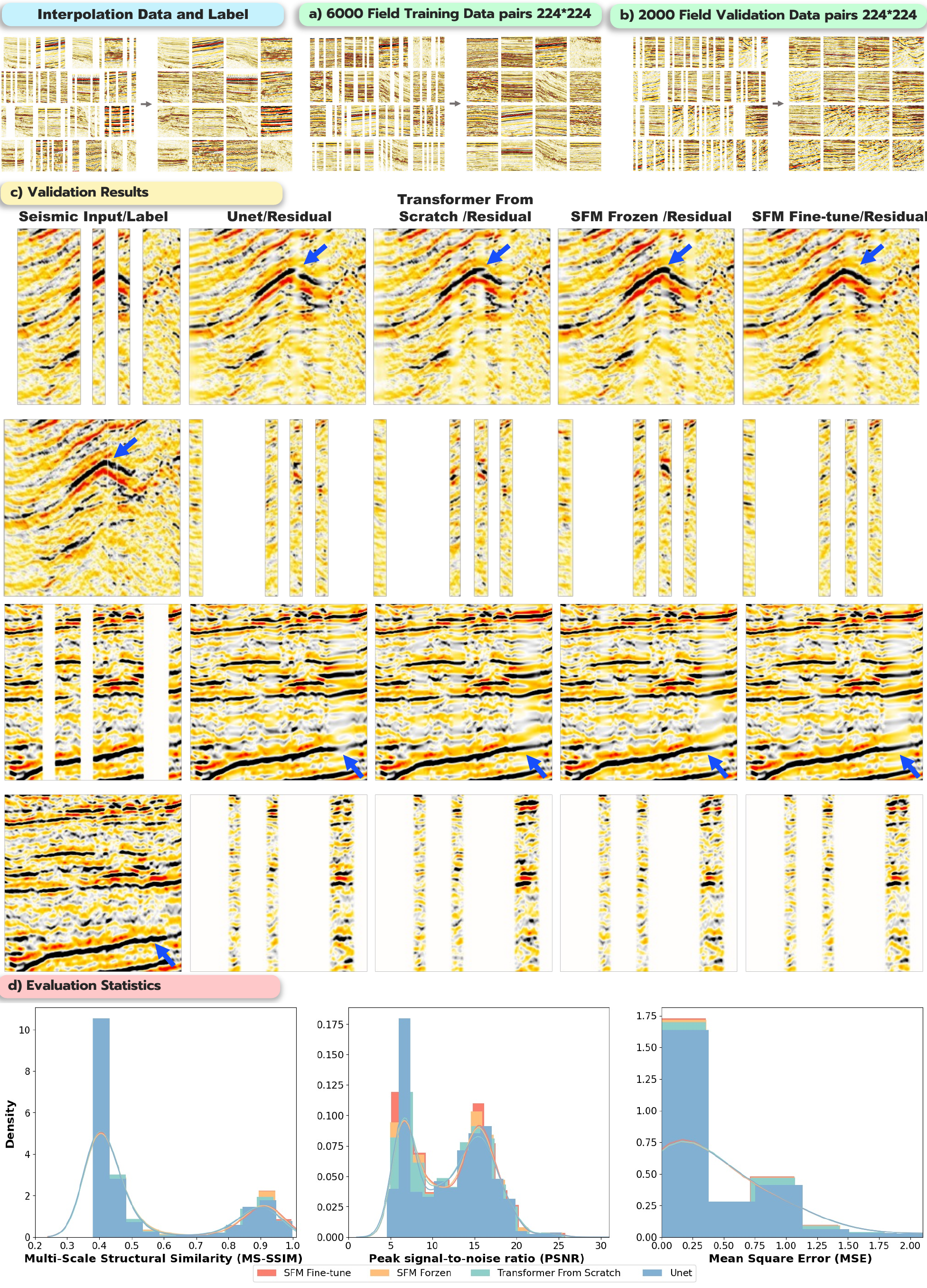}
\caption{
Seismic Interpolation Task: We employed 6000 field data for training and 2000 field data for validation. For this task, we conducted comparative experiments on the models of the Unet, Transformer From Scratch, SFM Frozen, and SFM Fine-tune. Predictions from the Unet mode reveal discontinuities in interpolated structures. Comparatively, predictions from the other three models tend to maintain a more continuous structural representation. The statistical graphs showcase MSE, PSNR, and MS-SSIM between predictions and original data. Notably, predictions from the SFM Frozen and SFM Fine-tune models tend to align more in regions of higher structural similarity, higher PSNR, and lower MSE. This implies that our predicted results closely resemble the original data in both structural and numerical aspects.
}\label{fig6}
\end{figure}

In the field of geophysics, another important problem that requires attention is interpolation. During the process of signal acquisition, the observation system may be limited by natural and human factors, leading to data missing in some areas. Interpolating the missing data is beneficial for data analysis and subsequent inversions. Seismic interpretation can also help geophysicists and geologists interpret geological structures, describe stratigraphic contacts, and gain a more comprehensive understanding of subsurface formations. Interpolation algorithms have been developed and widely applied by many experts, and traditional methods work well for cases with a low proportion of missing and discontinuous traces. However, for situations with a large proportion of missing and continuous traces, the problem becomes more challenging. In this context, we selected 6000 seismic data from field as the interpolation training dataset and randomly masked a portion of continuous seismic traces, setting the missing ratio at 25\% (Fig. \ref{fig6}a). The corresponding labels for the input data were known because we possessed complete seismic data. Similarly, we selected 4000 field seismic data samples for building the validation dataset (Fig. \ref{fig6}b).

For the interpolation task, we also compared the performance of the Unet, Transformer From Scratch, SFM Frozen, and SFM Fine-tune models. Fig. \ref{fig6}c displayed the interpolation results and residuals, where we can observe that the Unet model tended to disconnect continuous geological structures. In comparison, SFM Frozen and SFM Fine-tune methods demonstrated better continuity for complex geological structures. We employed Mean Square Error (MSE), MS-SSIM, and Peak Signal-to-Noise Ratio (PSNR) to evaluate the similarity between the interpolated results and the labels (Fig. \ref{fig6}d). From the distribution on the validation datasets and the overall average metrics (Table \ref{tab2}), we can observe that SFM Frozen and SFM Fine-tune models performed better. These results indicated that our foundation model can reasonably complete missing information by attending to global features, making the seismic data more visually reflective of subsurface geological structures.

\begin{table}[htb!]
\begin{center}
\begin{minipage}{\textwidth}
\caption{
Regression Task Performance
}\label{tab2}
\begin{tabular*}{\textwidth}{@{\extracolsep{\fill}}lcccccc@{\extracolsep{\fill}}}
\toprule
& \multicolumn{1}{@{}c@{}}{Inversion} & \multicolumn{2}{@{}c@{}}{Signal Processing} & \multicolumn{3}{@{}c@{}}{Interpolation} \\\cmidrule{2-2}\cmidrule{3-4}\cmidrule{5-7}%
Method & MSSSIM & MSSSIM & MSSSIM-R$\downarrow$ &   MSE & MSSSIM &   PSNR  \\
\midrule
Unet  									& 0.4105 & 0.6964  & 0.6259 & 0.7416& 0.5376 & 12.1442\\
 \tabincell{l}{Transformer \\From Scratch}  			& 0.4073 & 0.9131  & 0.5743 & 0.7235& 0.5377 & 12.2200\\
\midrule
SFM Frozen            & 0.4148 & 0.9280 & \textbf{0.4181} &  0.7197& \textbf{0.5414} & 12.2569\\
SFM Fine-tune         & \textbf{0.4383} & \textbf{0.9547}  & 0.4936 & \textbf{0.7108}& \textbf{0.5414} & \textbf{12.2922}\\
\bottomrule
\end{tabular*}
\footnotetext{
Note: Bold values represent the best performance. (R) represent the residual noise}
\end{minipage}
\end{center}
\end{table}

\section{Discussion}
In this article, we presented an entire workflow for developing a seismic foundation model (SFM) and collected datasets for pre-training and evaluation. The worldwide sourced data from 192 3-D seismic datasets constituted a feature-rich dataset. This feature-rich dataset provided ample resources to train our robust feature extraction model. Leveraging the pre-trained SFM for feature extraction from seismic data proved highly effective in addressing various geophysical challenges, including classification (i.e. seismic facies), segmentation (i.e. seismic geobody), inversion (i.e. reflectivity estimation), signal processing (i.e. denoise), and interpolation. Beyond offering our pre-trained SFM, we provided the pre-training dataset as well as downstream task training and validation datasets for model evaluation. These datasets could prove beneficial in overcoming the prevalent challenge of evaluating deep learning methods across various data sources. As most current deep learning methods rely on individually collected or synthesized datasets for validation and aid in addressing specific regional problems, they fall short of accurately assessing the true impact of deep learning, potentially leading to misguided enthusiasm. Hopefully, the data we have collected will provide a fair comparison platform for evaluating models.

To verify the effectiveness of our pre-trained model, we conducted a comprehensive assessment of SFM across various tasks. The comparative analysis with Transformer From Scratch served to underscore the substantial enhancements conferred by pre-trained parameters. SFM Frozen's performance underscored its capacity to extract valuable features that guide downstream tasks, affirming the SFM's adeptness at discerning vital seismic data insights. Further, our experiments with SFM Fine-tune, involving overall fine-tuning, manifested superior results in contrast to the practice of freezing all Encoder parameters during post-training. The commendable outcomes of both SFM Fine-tune and SFM Frozen can be attributed to two critical factors. Firstly, the robust architecture of the Transformer framework plays an instrumental role. This advantage can be attributed to the Transformer's global attention mechanism, which allows the extraction of global seismic features for the various seismic processing and interpretation tasks. Secondly, initializing the foundation model with enhanced parameters empowered the network to capture pivotal seismic features more effectively. Our SFM Frozen and SFM Fine-tune models exhibited substantial performance gains when compared to the same architecture without the benefit of pre-trained foundation model parameters (Transformer From Scratch). This substantiated that a large and diverse unlabeled dataset significantly contributed to a proficient Encoder, essential for deciphering seismic data complexities, aligning seamlessly with the SFM's intended utility and performance.

 The versatility of our SFM was evident through its superior generalization observed in seismic facies classification and seismic geobody identification tasks. This attribute facilitated the expansion of our model's expertise across a broader spectrum of data by synergizing with limited expert-interpreted experiences, thereby presenting a potent solution to diverse geophysical challenges. On another facet, our SFM, fortified by extensive data-driven training, harmoniously integrates with the physics-driven element of synthetic data in regression tasks. This synergy culminates in an advanced capacity to resolve geophysical conundrums more effectively. The juxtaposition of empirical and synthetic data sources augments the model's overall problem-solving aptitude, exemplifying the comprehensive capabilities of our SFM model in addressing a diverse array of geophysical inquiries.

Our pre-trained seismic foundational model, which demonstrated stronger versatility, better generalization, scalability, a greater attention on global information, and full utilization of vast unlabeled data, aligned well with our expectations for seismic models. \\
1. \textbf{Multi-task versatility}: Through testing on multiple tasks, our model exhibited good performance across various tasks. \\
2. \textbf{Generalization}: Our model made stable predictions on validation data that are significantly different from the training data, indicating strong generalization. \\
3. \textbf{Scalability}: Our foundation model pre-trained on 224$\times$224 data can also be successfully applied to 768$\times$768 data in the downstream tasks. \\
4. \textbf{Global information Attention}: Leveraging the Transformer architecture enabled our model to capture global information, as demonstrated in multiple tasks (seismic facies classification, seismic geobody identification, reflectivity model inversion, denoising, and interpolation). Our method excels in achieving high structural continuity, a key aspect we desire in practical applications of neural networks.\\
5. \textbf{Full utilization of vast unlabeled data}: Through self-supervised pre-training on massive amounts of unlabeled dataset of field seismic data, our seismic foundation model can effectively alleviate the lack of labeled training data in practical deployment of deep learning models. With its superior feature extraction capabilities, the foundation model can be efficiently and effectively fine-tuned for a specific downstream task through transfer-learning, even with on a small training dataset (e.g., 100 seismic facies training pairs) for that particular task. This presents a potential application: we can interpret a small portion of profiles with expert knowledge, and the foundational model can then generalize this knowledge to the rest of uninterpreted data in practical production.


In the application of our SFM, we found that there are still some issues related to the foundation model that need to be explored, such as how to utilize the computer vision pre-trained model to help geophysical problems and how to choose the Decoder when applying our model. While leveraging pre-trained computer vision models may seem convenient, effectively matching seismic data to natural images to make computer vision models adaptable to seismic images is an area that requires further exploration. We tested a simple approach by replicating seismic data three times to create three-channel data, then fine-tuning a pre-trained computer vision model on downstream tasks. However, due to the differences between seismic data and natural images, this approach did not perform as well as our foundational model. However, it is still a potential direction if we find a way to map seismic data to natural images.

Another issue worth exploring is about Decoder architecture, we found that fusing outputs from multiple layers was beneficial for segmentation tasks. The shallow layers of the Transformer block in Vision Transformer focus on local information, while the deeper layers emphasize global information. Fusing both local and global information is more advantageous for segmentation tasks (Supplementary Fig. \ref{fig:supp_figure1}). For regression tasks, a more complex Decoder is required, thus we introduced more learnable parameters and adopted a Unet-like module (like the right expansion path of Unet) for continuous upsampling in the Decoder (Supplementary Fig. \ref{fig:supp_figure1}). It's worth noting that the Decoder architecture we used is illustrative, and exploring how to leverage the foundational model's Decoder architecture for different tasks deserves further research. More details about the Decoder architecture are included in the supplementary file.

In fact, the training approach presented in this paper can be applied to train other foundation models for geoscience. Here are some feasible directions for the future: \\
1. \textbf{Professional Geoscience Language Models based on language models and knowledge graphs}: Utilizing geoscience literature and books, these models can provide professional knowledge in a question-answer format, promoting the dissemination of geoscience knowledge. \\
2. \textbf{Multimodal Geoscience Models}: Training language models to guide image models is an emerging trend. By combining language with geoscience data, we can train models more efficiently (leveraging pre-trained language models). Language can serve as a prompting engine, defining or guiding geoscience tasks (e.g., how to classify seismic data into facies) and generating seismic science schematic diagrams (e.g., schematic diagrams of normal faults, salt domes, and subsurface structures). \\
3. \textbf{Lightweight Deployment of Large Geoscience Models}: Future directions include knowledge distillation to reduce model memory consumption while preserving performance. This allows for faster inference, enabling real-time monitoring, disaster risk warning, and other applications. Moreover, lightweight models combined with interaction allow experts to continuously correct model results. For instance, experts can provide 1-2 points or rough lines to interpret the distribution of faults/salt domes in the subsurface or interpret seismic facies. Alternatively, they can input the necessary corrections to the model using text.\\

In conclusion, we believe that foundation models in geoscience are a feasible and promising trend for the future. They serve as a breakthrough in the application of AI in the field of intelligent geoscience. By harnessing the power of the powerful feature extraction capabilities of the foundation models, they can offer superior versatility, better generalization, and enhanced performance in various geophysical tasks, opening new possibilities for advancing geoscience data analysis, subsurface structure interpretation, and other geoscience investigations.

\section{Methods}\label{sec4}

The application of our foundation model (as shown in Supplementary Fig. \ref{fig:supp_figure1}) involved two stages of training: Pre-training and Downstream task application. In the Pre-training stage (upper part of Supplementary Fig. \ref{fig:supp_figure1}), we used a diverse set of training dataset and employed the Masked Autoencoders ($MAE$) method \citep{He_2022_CVPR} as the Pre-training method. In the Downstream task stage, we utilized the parameters obtained from our pre-trained foundation model to extract features, which were then fed into a simple Decoder network for further Downstream task applications. The Decoder network was divided into two categories of segmentation (seismic facies classification and seismic geobody identification) and regression tasks (denoising, reflectivity model inversion, and interpolation). Although both tasks involve pixel-level outputs, their levels of difficulty differ. It is necessary to design different types of Decoder networks tailored to optimize the model's performance on different tasks.

\subsection{Pre-train Stage}\label{subsec5}
1. \textbf{Embedding Module}: The input seismic data was divided into non-overlapping patches of size 16$\times$16. 75\% of these patches were randomly masked, and the remaining 25\% were fed into the network to obtain patch features through a learnable linear projection (patch embedding). We also incorporated 2-D position embedding to retain positional information. Here, we compared the mask ratio of 50\% and 75\%. Experiments showed that with 20\% additional information provided, the reconstructed image is improved by 6.3\% on MS-SSIM, MSE is reduced by 15.6\%, and PSNR is improved by 4.0\%. Masking 50\% of the patches resulted in better image restoration, but our goal is to train a better Encoder that can reconstruct images with fewer visible patches. Therefore, we chose a mask ratio of 75\%. When the ratio was set to 90\%, the input information would be too sparse (e.g., in a 224$\times$224 image, only 20 patches are visible to predict 196 patches), making training hard to converge and preventing the attainment of a high-quality Encoder.

2. \textbf{Encoder Module}: After obtaining the embedding for each patch, we fed these embeddings into multiple Transformer Blocks. Each Transformer block consisted of a MSA (Multiple Self-Attention) module and a MLP (Multi-Layer Perceptron) module. Layer Norm was applied to the features before feeding them into the modules to stabilize the distribution of feature values. Besides, Residual Connection was applied after the modules' output to enhance the network's ability to overcome gradient vanishing. Here, we provided two Encoder modules to be adapted to different tasks: ViT-Base and ViT-Large. The ViT-Base model consists of 12 Transformer Layers, a Hidden size (D) of 768, an MLP size of 3072, and 12 heads. The ViT-Large model comprises 24 Transformer Layers, a Hidden size (D) of 1024, an MLP size of 4096, and 16 heads.

3. \textbf{Decoder Module}: We merged the features obtained from the Encoder's output with the learnable features of the masked patches. These merged features were then collectively fed into the Decoder's Transformer Block to produce the final features. Considering that seismic images were less complex than natural images, we performed simple tests on a small dataset to strike a balance between computational efficiency and accuracy (Supplementary Table \ref{tab4}). As a result, we decided on the following Decoder configuration: 4 Transformer Layers, a Hidden size (D) of 256, an MLP size of 1024, and 16 heads.

4. \textbf{Output Module}: The output features were transformed into pixel features through a linear projection and subsequently reshaped into seismic images.

\subsection{Downstream Task Application Stage}\label{subsec6}

The well-trained foundation model obtained by the Pre-training Stage was utilized for feature extraction in downstream tasks. We employed the extracted features as input to a simple Decoder. Two types of Decoders were designed to accommodate different task complexities: one for segmentation tasks (seismic facies classification and seismic geobody identification) and the other for regression tasks (denoising, reflectivity model inversion, and interpolation). For the Decoder used in segmentation tasks, we adopted a structure similar to SERT-MLA\citep{Zheng_2021_CVPR}. Specifically, we concatenated the outputs from the 3rd, 6th, 9th, and 12th (or 6th, 12th, 18th, and 24th for the Large model) Transformer blocks of the Base (or Large) foundation model. We then performed 4x upsampling and applied a set of convolutional layers. To match the size of the input image, we also implement an upsampling operator.

As for the Decoder used in regression tasks, we adopted a similar upsampling module to Unet. We took the output from the last layer of our foundation model and fed it through a series of operations: convolution, 2x upsampling, convolution, repeated four times to maintain the feature dimension consistent with that of the input data. Additionally, we applied a convolutional operation to the input data and concatenated it with the output of the upsampling module. Finally, we used a convolutional module to produce the ultimate regression result. 

\subsection{Loss function and Evaluation metrics}\label{subsec7}

In the Pre-training stage, we chose Mean Squared Error (MSE) Loss as the Loss function. While computing the loss for masked patches, we did not apply mean-variance normalization to each small patch. Using normalization weakens the amplitude contrast information, which distinguishes seismic data from natural images. Therefore, we decided not to include patch mean-variance normalization when calculating the loss ( Supplementary Table \ref{tab4}). In the Downstream Task Application Stage, we used two different losses for segmentation and regression tasks, respectively. The choice of different loss functions for specific tasks was based on their respective requirements. For segmentation tasks (seismic facies classification and seismic geobody identification), the CrossEntropy loss is employed to measure the total entropy between the predicted output probability distribution and the label probability distribution. For regression tasks (denoising, reflectivity model inversion, and interpolation), the loss function consisted of a weighted combination of the MSE loss and the Multi-Scale Structural Similarity (MS-SSIM) loss. By incorporating the MS-SSIM loss, the network can learn more structural information. In addition, we calculated loss only in the missing part of the interpolation task. More training Settings are shown in the Supplementary Table \ref{tab3}.

Regarding the evaluation metrics, segmentation tasks (seismic facies classification, geological target detection) were assessed using Accuracy and Intersection over Union (IoU). For regression tasks (denoising, reflectivity model inversion, and interpolation), the evaluation metrics including MSE, PSNR, and MS-SSIM were employed. These evaluation metrics provided comprehensive measures to assess the performance and accuracy of the model in various geophysical applications, ensuring a thorough evaluation of the model's capabilities across different tasks.

\section{Data availability}
Pre-training and downstream task application data links are organized on \href{https://github.com/shenghanlin/SeismicFoundationModel}{GitHub}.
\section{Code availability}
Pre-training and downstream task application code links are organized on \href{https://github.com/shenghanlin/SeismicFoundationModel}{GitHub}. 


\bibliographystyle{plainnat}
%





\clearpage

\section{Supplementary Information}\label{sec7}

\setcounter{figure}{0}
\renewcommand{\thefigure}{S\arabic{figure}}
\begin{figure}[htbp!]%
\centering
\includegraphics[width=\textwidth]{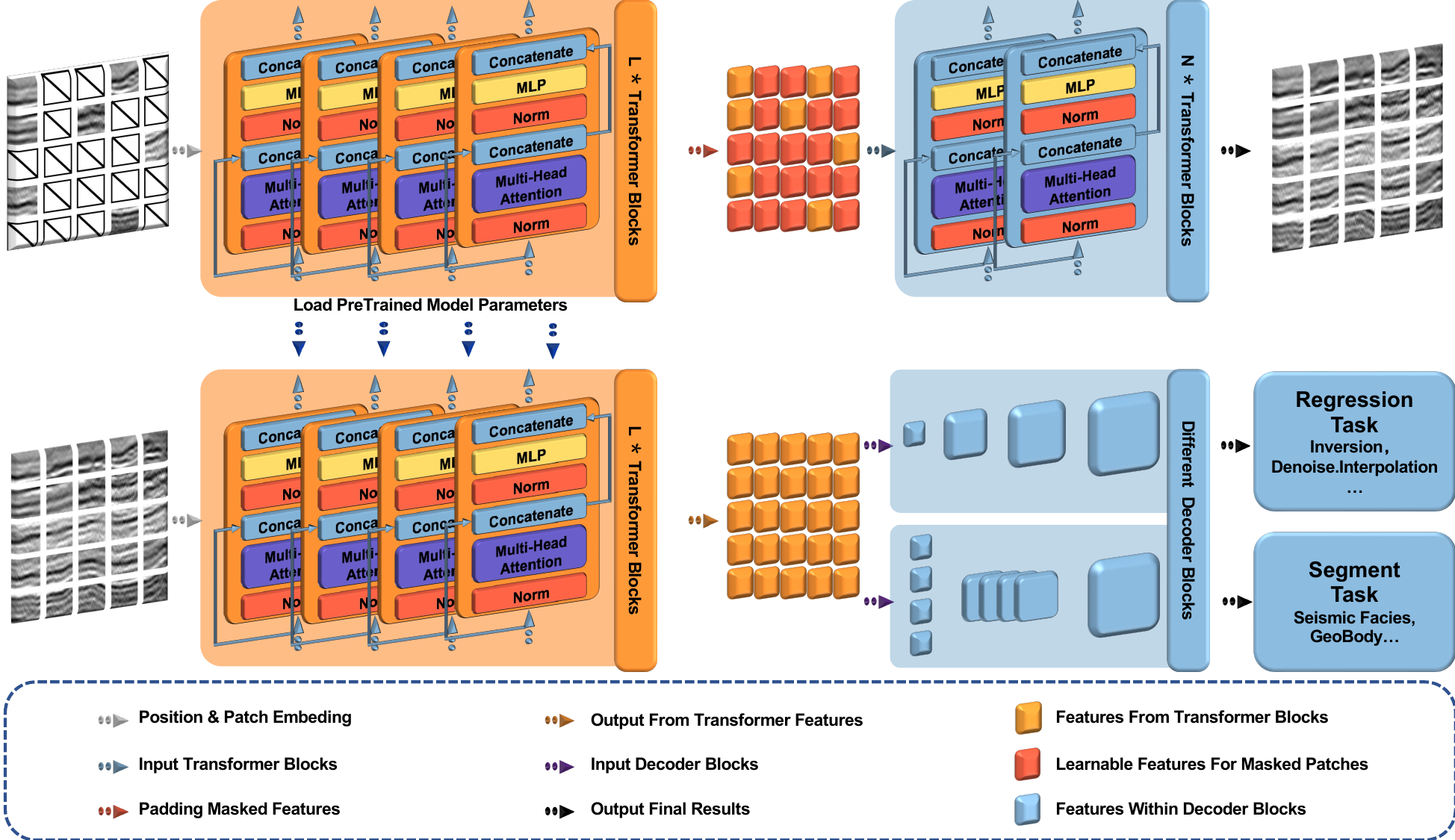}
\caption{Network architecture of Seismic Foundation Model (SFM) at two stages: Pre-training and Downstream Task Application. In the Pre-training stage, we utilize a diverse training dataset and employ the Masked Autoencoders (MAE) for initial feature extraction. Specifically, 75\% of randomly selected patch data undergo masking, followed by input into the encoder network. Subsequently, the missing features are supplemented with learnable features, forming a complete set of corresponding features. The Pre-trained decoder network then generates outputs for all patches. The trained encoder is repurposed as the Seismic Foundation Model for downstream tasks. In the Downstream Task stage, we leverage the SFM to extract features, which are subsequently fed into a simplified decoder network for further application. This downstream decoder network is bifurcated into two branches, each tailored to accommodate specific tasks such as segmentation, regression, and so on. For segmentation tasks, outputs are derived by concatenating outputs from various layers of Transformer Blocks, followed by interpolation of feature sizes to achieve the final output. For regression tasks, the final layer's output is progressively transformed through convolution-upsampling-convolution modules to match the input image's dimensions. Ultimately, these outputs are concatenated with specific features derived from convolutions applied to the input image.
}\label{fig:supp_figure1}
\end{figure}

\begin{figure}[htbp!]%
\centering
\includegraphics[width=\textwidth]{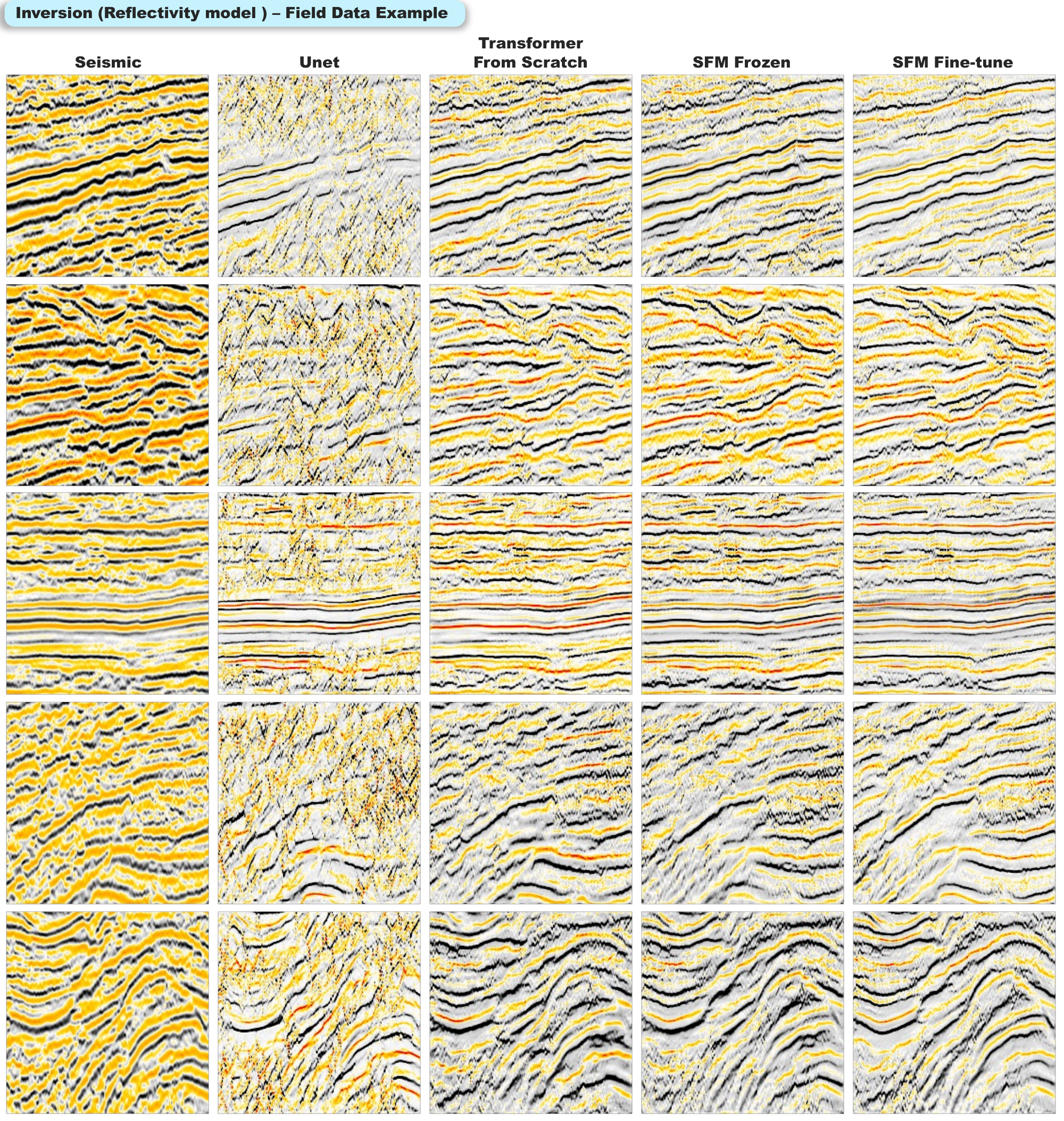}
\caption{Seismic Inversion (Reflectivity Model) Task on Field Data: We compared the performance of the Unet, Transformer From Scratch, SFM Frozen, and SFM Fine-tune models on field data. Despite the lack of labeling of the field data, we can still observe that the predictions of the Unet model exhibit erratic features in low signal-to-noise ratio areas and lack continuity in reflectors. Conversely, other three models displayed more continuous reflectors.}\label{fig:supp_figure3}
\end{figure}

\setcounter{table}{0}
\renewcommand{\thetable}{S\arabic{table}}
\begin{table}[htb!]
\begin{center}
\begin{minipage}{\textwidth}
\caption{
Ablation experiments on the decoder at the pre-training stage.
}\label{tab4}
\begin{tabular*}{\textwidth}{@{\extracolsep{\fill}}lccccccccc@{\extracolsep{\fill}}}
\toprule
Index & \multicolumn{2}{@{}c@{}}{Mask Ratio} & \multicolumn{2}{@{}c@{}}{Dim} & \multicolumn{3}{@{}c@{}}{Depth} & \multicolumn{2}{@{}c@{}}{Loss Function} \\\cmidrule{2-3}\cmidrule{4-5}\cmidrule{6-8}\cmidrule{9-10}%
 & 0.75 & 0.5 & 512 & 256 & 8& 4 & 1 & \tabincell{c}{w/o \\norm} & \tabincell{c}{w/ \\norm}  \\
\cmidrule{1-1}\cmidrule{2-3}\cmidrule{4-5}\cmidrule{6-8}\cmidrule{9-10}%
MS-SSIM  & 0.392  & \textbf{0.417}  & \textbf{0.392 } & 0.382  & \textbf{0.487}  & 0.480  & 0.462  & \textbf{0.392}  & 0.346\\
MSE         &  0.833  & \textbf{0.702}  & 0.833  & \textbf{0.799}  & \textbf{0.445}  & \textbf{0.455}  & 0.463  & \textbf{0.833}  & 0.947\\
PSNR       & 21.77  & \textbf{22.65}  & 21.77  & \textbf{21.93}  & \textbf{23.19}  & 23.10  & 23.01  & \textbf{21.77}  & 21.22\\
\bottomrule
\end{tabular*}
\end{minipage}
\end{center}
\end{table}

\begin{table}[htb!]
\begin{center}
\begin{minipage}{\textwidth}
\caption{
Pre-Training and Fine-tune Setting.
}\label{tab3}
\begin{tabular*}{\textwidth}{@{\extracolsep{\fill}}lcccccccc@{\extracolsep{\fill}}}
\toprule
Task & optimizer & base lr &  \tabincell{c}{batch \\size} &  \tabincell{c}{lr \\schedule} &  \tabincell{c}{warmup \\epochs} &  \tabincell{c}{training \\epochs}\\
\midrule
Pre-trained  & AdamW & 1.5e-4 &  9280 &  cosine &  40 &  1600 \\
\midrule
 \tabincell{c}{Facies \\Classifcation} & AdamW & 1.5e-3 & 1  & cosine  & 10 & 100\\
 \\
 \tabincell{c}{GeoBodies \\Identification}    & AdamW & 1.5e-3 & 64  & cosine  & 10 & 100 \\
 \\
 \tabincell{c}{Reflectivity \\Estimation}   & AdamW &  6.4e-4 & 60 & cosine & 10 & 100 \\
 \\
Denoise  &AdamW&  6.4e-4 & 60 & cosine & 10 & 100 \\
\\
Interpolation  &AdamW&  1.5e-4 & 50 & cosine & 10 & 300  \\
\bottomrule
\end{tabular*}
\footnotetext{
Note: AdamW\citep{loshchilov2017decoupled}, warmup epochs \citep{goyal2017accurate},lr schedule \citep{loshchilov2016sgdr}. lr represent learning rate. }
\end{minipage}
\end{center}
\end{table}
\clearpage

\end{document}